\documentclass[5p, authoryear]{elsarticle}
\usepackage{epsfig,subfigure,amsmath,graphicx}
\usepackage{color,hyperref}

\definecolor{purple}{rgb}{1,0,1}

\newcommand{\thetacut}{\theta_{\rm cut}}
\newcommand{\lcut}{\ell_{\rm cut}}
\newcommand{\lmax}{\ell_{\rm max}}
\newcommand{\zcut}{z_{\rm cut}}
\newcommand{\tgpu}{t_{\tt ARKCoS}}
\newcommand{\tcpu}{t_{\tt libpsht}}

\bibpunct[; ]{(}{)}{;}{a}{}{,}

\begin{document}

\begin{frontmatter}

\title{Using hybrid GPU/CPU kernel splitting to accelerate spherical convolutions}

\author[add1,add2,add3,add4,add5,add6]{P.~M.~Sutter}
\ead{sutter@oats.inaf.it}
\author[add3,add4,add6,add7]{Benjamin D. Wandelt}
\author[add3,add4,add8]{Franz Elsner}

\address[add1]{INFN - National Institute for Nuclear Physics, via Valerio 2, I-34127 Trieste, Italy}
\address[add2]{INAF - Osservatorio Astronomico di Trieste, via Tiepolo 11, 1-34143 Trieste, Italy}
\address[add3]{Sorbonne Universit\'{e}s, UPMC Univ Paris 06, UMR7095, Institut d'Astrophysique de Paris, F-75014, Paris, France}
\address[add4]{CNRS, UMR7095, Institut d'Astrophysique de Paris, F-75014, Paris, France}
\address[add5]{Center for Cosmology and AstroParticle Physics, Ohio State University, Columbus, OH 43210, USA}
\address[add6]{Department of Physics, University of Illinois at Urbana-Champaign, Urbana, IL 61801, USA}
\address[add7]{Department of Astronomy, University of Illinois at Urbana-Champaign, Urbana, IL 61801, USA}
\address[add8]{Department of Physics and Astronomy, University College London, London WC1E 6BT, U.K.}

\begin{abstract}
We present a general method for accelerating by more than an order of magnitude 
the convolution of pixelated functions on the sphere with a radially-symmetric kernel.  
Our method  splits  the kernel into a compact real-space component 
and a compact spherical harmonic space component.
These components can then be convolved in parallel using an inexpensive commodity GPU and a CPU. 
We provide models for the computational cost of both real-space and Fourier space convolutions and an estimate for the approximation error.
Using these models we can determine the 
optimum split that minimizes the wall clock time for the convolution while satisfying the desired error bounds. 
We apply this technique to the problem of simulating a cosmic microwave 
background (CMB) anisotropy 
sky map at the resolution typical of the high resolution maps 
produced by the 
Planck mission. For the main Planck CMB science channels we 
achieve a speedup of over a factor of ten, assuming an acceptable fractional rms error of order $10^{-5}$ in the power spectrum of the output map.
\end{abstract}


\end{frontmatter}

\section{Introduction}
\label{sec:introduction}

Current and next-generation cosmic microwave background (CMB) experiments, 
such as Planck \linebreak \citep{PlanckCollaboration2011},
the Atacama Cosmology Telescope \citep{KOSOWSKY2003},
the South Pole Telescope \citep{Ruhl2004},
and CMBPol \citep{Baumann2009}
promise a great wealth of cosmological 
and astrophysical information \citep{Smoot2010}.
The most common operation in CMB data analysis consists of convolving  a real or synthetic map with a radial kernel.
Large numbers of such  smoothing or filtering operations are necessary for 
many critical data analysis applications, such as the simulation of 
CMB maps \citep{Gorski2005},
map-making from multichannel 
maps \citep{Tegmark1997, Natoli2001, Stompor2001, Patanchon2008, Sutton2010}, 
iterative calculations of inverse covariance weighted data (e.g., in the 
context of optimal power spectrum estimation or 
Wiener filtering \citep{Wandelt2004}),
wavelet analysis \citep{Hobson1999, Martinez-Gonzalez2002, Vielva2004}, 
point-source removal \citep{Tegmark1998, Gonzalez-Nuevo2006},
and destriping errors induced by noise removal \citep{Efstathiou2007}.

Outside of CMB analysis, 
the future Euclid mission~\citep{Laureiji2011} will resolve the 
sky to sub-arcsecond resolution, and one technique for identifying 
overdensities in such a map is via convolution with a filter.
In addition, a variety of fields require the 
regular use of spherical convolution operations in certain applications, 
such as in
geophysics and meteorology~\citep{Vanicek2003},
medical imaging~\citep{Yeo2008}, and
computer vision and environment simulation~\citep{Miller1984}.

Until recently, the near-exclusive practice in the CMB community to compute radial kernel convolutions was to use the spherical convolution theorem: 
transform to spherical harmonic space, multiply 
the spherical harmonic coefficients with the $\ell$-space 
representation of the radial kernel, and back-transform to pixel space. 
As a consequence of the ubiquity of radial kernel convolution for data analysis on the sphere and the ready availability of software implementing the discrete forward and backward fast Spherical Harmonic Transformation (SHTs), this has become the major application for SHTs. 
Interest in the actual $a_{lm}$ coefficients is relatively rare by comparison.

Graphics Processing Units (GPUs) offer a promising solution 
to the  computational challenges posed by radial kernel convolution 
to current and upcoming 
data sets on the sphere~\citep{Brunner2007, Barsdell2010, Fluke2011}  due to their low cost and high degree of parallelism.
Indeed, the recent rise of 
cheap GPU hardware and associated extensive programming 
libraries have led to their use in
many applications in astrophysics, such as
the analysis of the Lyman-$\alpha$ forest~\citep{Greig2011},
dust temperature calculations~\citep{Jonsson2010},
magnetohydrodynamics~\citep{Pang2010},
adaptive-mesh refinement simulations~\citep{Schive2010},
analysis of data from the Murchison Widefield Array~\citep{Wayth2007},
volume renderings of spectral data from the Australian 
Square Kilometer Array Pathfinder mission~\citep{Hassan2011},
and visualizations of large-scale data sets~\citep{Szalay2008}.

While GPU implementations of the SHT \citep{HUPCAIoanO.2010, Szydlarski2011} 
have only achieved modest speed-ups, ~\citet[][hereafter EW11]{Elsner2011} tackled the problem of spherical 
convolutions for compact radial kernels by specifically designing an algorithm  adapted to benefit from the high degree of parallelism and memory bandwidth for compact kernels.
Compared to the serial time of a highly optimized implementation of the Fast SHT algorithm \cite{Reinecke2011}, EW11 demonstrated a speed-up 
of up to a factor of 
60 using a commodity GPU costing \$500 with the further benefit of strongly suppressing Fourier ringing artifacts.  Other approaches, 
such as optimizing traditional algorithms~\citep{Muciaccia1997}
and using large-scale computing resources~\citep{GhellerC.2007}, either 
do not scale as efficiently or do not exploit readily available hardware.
The main limitation of the method described by EW11 is that significant speed-ups can only be achieved for relatively compact kernels. 
While there are still many applications for such compact kernels, 
such compactness can lead to unreasonable artefacts in the 
resulting smoothed maps.

To take advantage of the power of GPUs with kernels of arbitrary size, 
we must split the given kernel between a real-space portion, which will 
be applied using a GPU, and an $\ell$-space (i.e., Fourier) 
portion, which will be applied 
using traditional CPU methods. Each portion of the full kernel will then 
necessarily be truncated, resulting in a small --- but predictable --- error. 
Given an upper bound for an acceptable error, we must determine the 
optimal splitting between real- and $\ell$-space in order to achieve 
maximum performance.

In this work we present 
{\tt scytale} \footnote{We take the name from the ancient cryptographic 
system where only rods of a precise radius could be used to decode messages.}, 
a tool for splitting kernels between truncated real- and $\ell$-space 
portions, estimating the errors due to the truncations,
 and discovering the optimum truncations for a given kernel.
We apply this tool to determine the expected speedup 
when splitting a given kernel 
between the GPU code {\tt ARKCoS} of EW11 and the CPU code 
{\tt libpsht} of~\citet{Reinecke2011}.
In Section~\ref{sec:optimization} we discuss our strategy for 
splitting kernels, estimating errors, and determining the optimum truncations. 
We present an analysis of the errors and our optimization results in 
Section~\ref{sec:results}, followed by a discussion and conclusion 
in Section~\ref{sec:conclusions}.

\section{Estimating Errors \& Optimization Strategy}
\label{sec:optimization}

We decompose a given kernel $K_\ell$ into truncated $\ell$-space and real-space 
portions, which we denote as $\widehat{K}_\ell$ and $\widehat{K}_\theta$, 
respectively. We may then construct an approximate kernel as
\begin{equation}
  \widetilde{K}_\ell = \widehat{K}_\ell + P_{\ell \theta} \widehat{K}_\theta,
\label{eq:approx}
\end{equation}
where $P_{\ell \theta}$ is a Legendre transformation operator.
We truncate the $\ell$-space kernel to a limit $\lcut$ and the 
real-space kernel to a limit $\thetacut$.

For a given $\lcut$ and $\thetacut$, we construct the functional 
forms of the truncated kernels by simultaneously minimizing 
the root-mean-square error of the $\ell$-space kernel,
\begin{equation}
  \sigma_{\rm rms}^2 \equiv 
   \frac{1}{4 \pi}
   \sum_{\ell=0}^{\lmax} 
   \left(
     \left( K_\ell - \widetilde{K}_\ell \right)^2 C_\ell^{\rm input} 
            (2 \ell + 1),
   \right)
\label{eq:minimize}
\end{equation}  
and a similar expression for $K_\theta$.
In the above, the input power spectrum 
$C_\ell^{\rm input}$ depends on the particular application; 
for simulating CMB maps which simply contain uncorrelated noise, 
this will be a constant, whereas for Wiener filtering this 
will have a spectrum $\sim C_\ell^2 / (C_\ell + N)$, where $N$ 
is the noise covariance.

We compute this weighted least-squares fit by solving the matrix-vector 
equation
\begin{equation}
  \mathbf{A}^\mathrm{T} \mathbf{W} \mathbf{A} x 
     = \mathbf{A}^\mathrm{T} \mathbf{W} K_\ell,
\label{eq:leastsquares}
\end{equation}
where our solution vector $x$ is a concatenation of the $\ell$- and 
real-space kernels:
\begin{equation}
  x_i = \left\{ 
       \begin{array}{rl}
       \widehat{K}_\ell   &\mbox{ $i \le \lcut $}\\
       \widehat{K}_\theta &\mbox{ otherwise}
       \end{array}
       \right.
\label{eq:solvec}
\end{equation}  
The vector representation of $\widehat{K}_\theta$ contains 
$\zcut$ elements, where $\zcut \equiv \lmax - \ell_{\cos{\thetacut}}$.
We similarly construct the matrix $\mathbf{A}$, which has 
$\lmax$ rows and $(\lcut+\zcut)$ columns, such that
\begin{equation}
  A_{i, j} = \left\{ 
       \begin{array}{rl}
       \delta_{i, j} &\mbox{ $j \le \lcut$}\\
       P_{i, j}      &\mbox{ otherwise},
       \end{array}
       \right.
\label{eq:matrix}
\end{equation}  
where $P_{i,j}$ are elements of the Legendre transformation operator.
Finally, the elements of the weighting matrix are given by 
$W_{i,j} = (2 i +1) \delta_{i, j}$.

The matrix $\mathbf{A}^\mathrm{T} \mathbf{W} \mathbf{A}$ is nearly 
degenerate and thus difficult to invert directly. Additionally, the 
nearly-degenerate modes add undesirable large-amplitude fluctuations to 
the final solution. Thus, we use 
standard singular value decomposition (SVD) techniques
to solve Eq.(\ref{eq:leastsquares}).
By neglecting any singular values below $10^{-6}$, we damp the large 
oscillations.
Once we have the truncated kernels, we can 
evaluate the resulting error by taking the fractional root mean square:
\begin{equation}
  \sigma^2 = \alpha^2 \frac{\sigma_{\rm rms}^2} 
                   {1 / 4 \pi \sum (2 \ell+1) K_\ell^2},
\label{eq:error}
\end{equation}
where the sum runs from 0 to $\lmax$.
The constant $\alpha$ represents any additional errors introduced 
by the actual convolution, such as those caused by 
single-precision arithmetic and inadequate kernel interpolation, 
and must be empirically determined.
Thus, given a particular kernel, this procedure allows us to 
identify values of $\lcut$ and $\thetacut$ that 
satisfy a given error bound.

If a particular $\lcut$ and $\thetacut$ satisfy an error bound, we 
then estimate the computational cost associated with the truncated kernels.
We assume the real-space portion will be solved using {\tt ARKCoS}
on a GPU, so we denote the cost as $\tgpu$. Similarly, 
we assume the $\ell$-space kernel will be solved using the standard 
library {\tt libpsht} on a CPU, 
and hence we will denote the cost as $\tcpu$.
The cost for applying the real-space GPU kernel is
\begin{equation}
  \tgpu = 0.0232{\rm s}~\thetacut + 2.428{\rm s} 
\label{eq:costGPU}
\end{equation}
and the cost for the $\ell$-space CPU kernel is
\begin{equation}
  \tcpu = 160{\rm s}~\frac{\lcut^2 \lmax}{4096^3}.
\label{eq:costCPU}
\end{equation}
Above, $\thetacut$ is in arcminutes.
To determine these scalings we used
an NVIDIA GeForce GTX 480 GPU and a 2.8 GHz Intel Core2 Quad CPU.
Our GPU scaling is different than the study of EW11 due to updated 
NVIDIA drivers.
Note that the CPU timing assumes the use of only a single core.
We assume throughout a data set with 
HEALPix~\citep{Gorski2005} 
resolution $n_{\rm side} = 2048$ and $\lmax=4096$, consistent 
with Planck observations~\citep{MennellaA.2011}.
Furthermore, we assume a power spectrum derived from WMAP 7-year 
results~\citep{Komatsu2011}.

\begin{figure} 
  \centering 
  {\includegraphics[type=png,ext=.png,read=.png,width=\columnwidth]{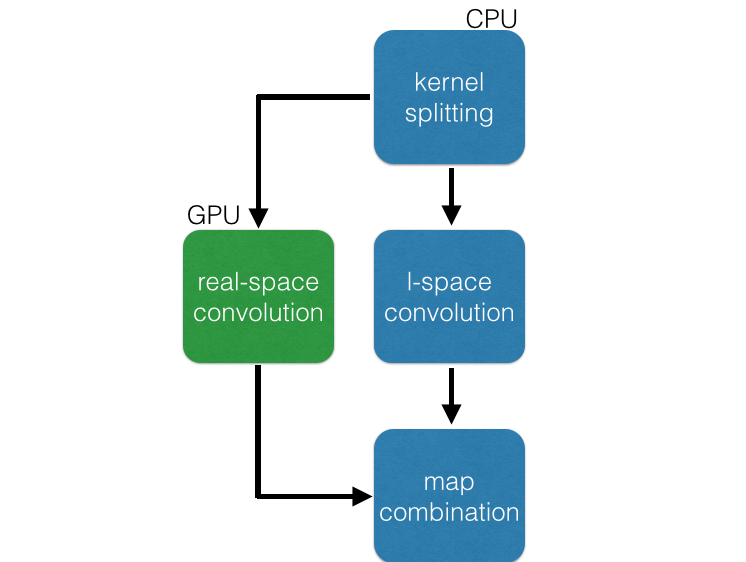}}
  \caption{Simple schematic showing the flow of computation. 
           The CPU performs the initial kernel splitting, then sends the 
           map and real-space kernel to the GPU. The CPU and GPU perform
           simultaneous convolutions with their respective kernels. 
           The CPU collects the real-space convolved map from the 
           GPU and adds it to its own map. The goal of optimization is 
           to minimize the difference in times between the two parallel
           convolutions.}
\label{fig:flowchart}
\end{figure}

We assume that the GPU and CPU portions can be solved in 
parallel, as shown in the simple flowchart diagram of 
Figure~\ref{fig:flowchart}, 
and hence our goal for a given kernel is to find the 
pair $(\lcut, \thetacut)$ that satisfies the error bound and 
at which $\tgpu=\tcpu$, minimizing the overall cost.
To find the optimum truncation we follow a straightforward 
scanning strategy of a linear search through values of $\thetacut$. 
For each $\thetacut$ we employ a
binary search in $\ell$-space for the smallest $\lcut$ that satisfies 
the error bound. We then select the most cost-effective pair from this set.

\section{Results}
\label{sec:results}

We study radially-symmetric kernels of the type
\begin{equation}
  K_\ell = \sqrt{C_\ell} B_\ell,
\label{eq:kernel}
\end{equation}
where $C_\ell$ is the expected power in the 
given $\ell$-space bin and $B_\ell$ is the 
Legendre transform of a beam. We assume an identical band limit of $\lmax$
for both the input power spectrum and the kernel. 
These particular kernels have a wide variety 
of applications. We assume a Gaussian beam 
with a given FWHM. For this analysis, we will also assume 
$C_\ell^{\rm input} \sim 1$ (that is, the case of simulating 
CMB maps with uncorrelated noise).

\begin{figure} 
  \centering 
  {\includegraphics[width=\columnwidth]{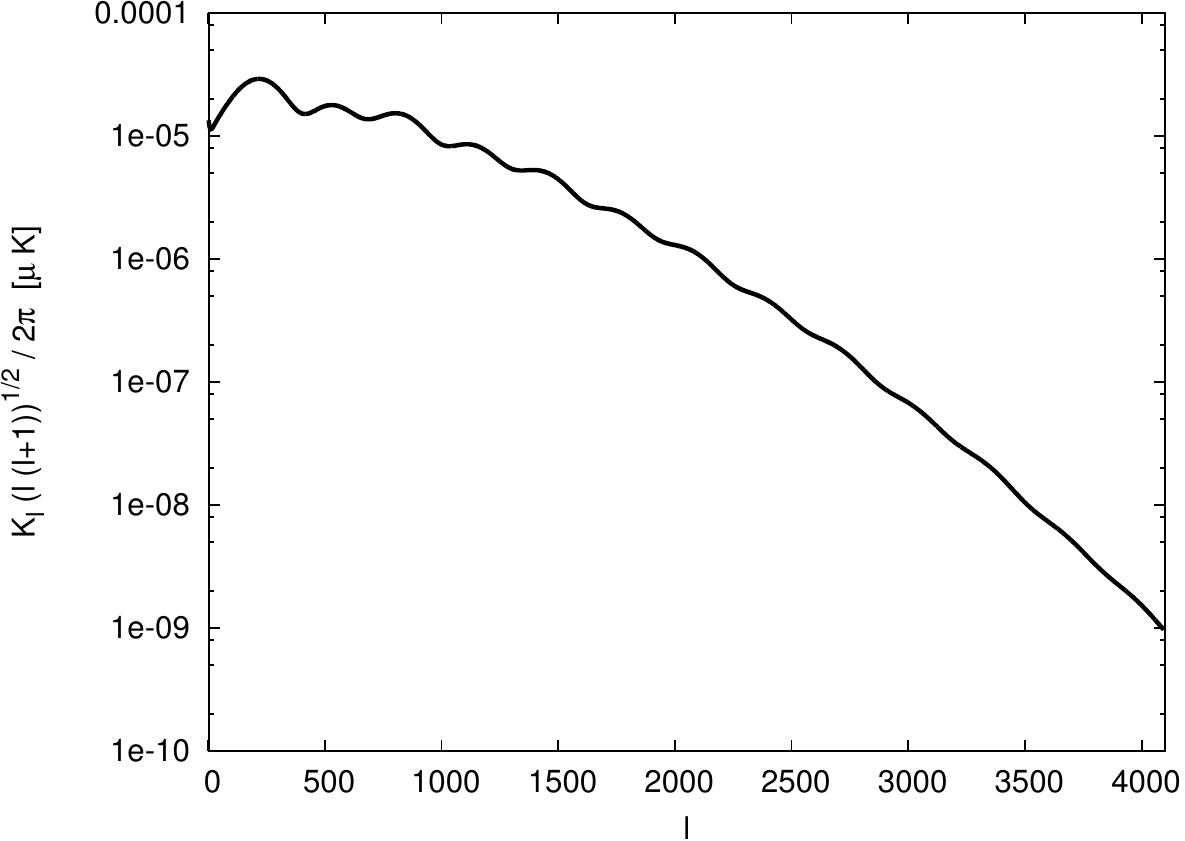}}
  \caption{Example kernel (Eq.~\ref{eq:kernel}) 
           for a beam with 7$'$ FWHM.}
\label{fig:kernel}
\end{figure}

We begin with an analysis of splitting a single kernel. 
We show in Figure~\ref{fig:kernel} an example kernel produced 
with a 7$'$ FWHM beam. We truncate the kernel and the input 
power spectrum at $\lmax=4096$. This narrow beam produces wide 
support to significantly high $\ell$: only past $\ell \approx 2000$ 
does the kernel drop below $1 \%$ of $\sqrt{C_\ell}$.

\begin{figure} 
  \centering 
  {\includegraphics[width=\columnwidth]{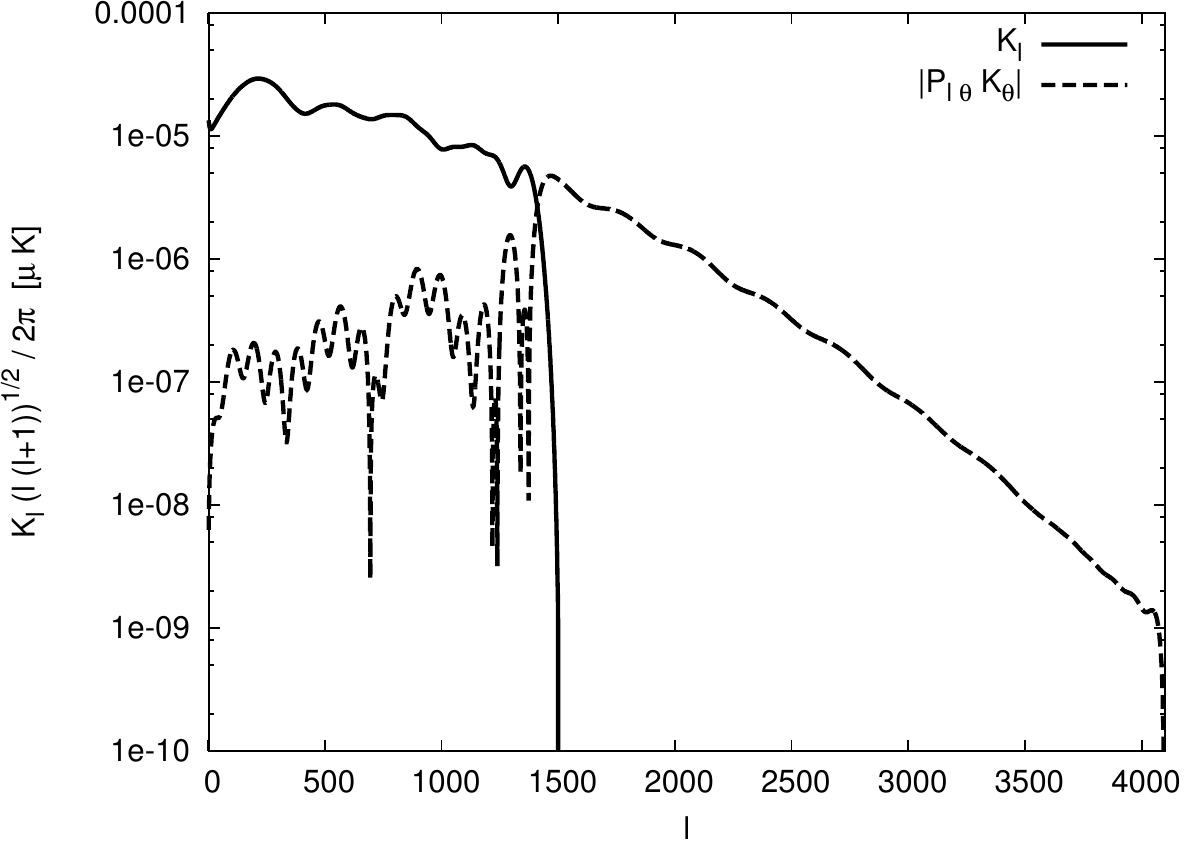}}
  \caption{Truncated $\ell$-space kernel (solid line) 
           and Legendre-transformed 
           truncated real-space kernel (dashed line) 
           for 
           the example input kernel with 7$'$ beam. 
           The $\ell$-space kernel is truncated to $\lcut=1500$ and the 
           real-space kernel to $\thetacut=240'$. To highlight the 
           oscillations, we plot the absolute value of the real-space 
           kernel.}
\label{fig:kerneltruncate}
\end{figure}

We show in Figure~\ref{fig:kerneltruncate} an example of the truncated 
kernels computed by {\tt scytale}. 
In this example the real-space kernel ($\widehat{K}_\theta$) is truncated at 
$\thetacut=240'$
and the $\ell$-space ($\widehat{K}_\ell$) kernel is 
truncated at $\lcut=1500$.
For clarity, we have plotted the absolute value of the real-space kernel.
As expected, the $\ell$-space kernel faithfully reproduces the low-$\ell$ 
portion of the full kernel while the real-space kernel matches the 
high-$\ell$ regime. In order to fit the behavior of the full kernel 
at high $\ell$, the real-space kernel produces large oscillations at 
low $\ell$, which are compensated by percent-level adjustments in 
the $\ell$-space kernel. Summed together, these kernels reproduce the 
full input kernel, except at the very highest $\ell$ where the low 
magnitudes make a full fit difficult.

\begin{figure} 
  \centering 
  {\includegraphics[width=\columnwidth]{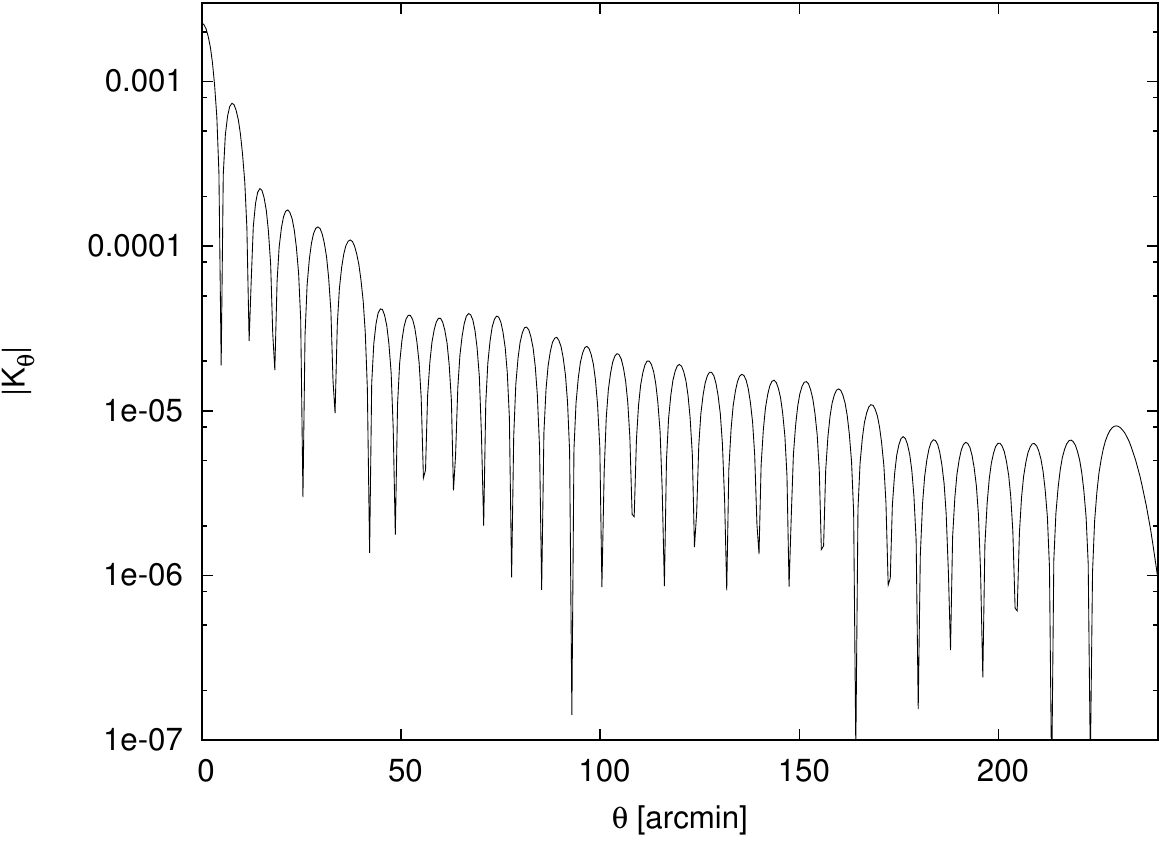}}
  \caption{Truncated real-space kernel for the example input kernel
           with 7$'$ beam. To highlight the oscillations, we plot 
           the absolute value of the kernel.}
\label{fig:kernelreal}
\end{figure}

Figure~\ref{fig:kernelreal} shows the truncated real-space kernel 
in real space itself. Even though our computational approach 
damps oscillations in $\ell$-space (where the fits to the 
full input kernel take place) we see rapid oscillations in the actual 
kernel that {\tt ARKCoS} uses in its real-space approach.
We must accurately 
interpolate this kernel, especially at small angles, 
in the convolution algorithm in order to both 
recover the high-$\ell$ behavior and correctly calculate 
the systematic offsets 
present in the low-$\ell$ portion of the approximate kernel.
To do this, we employ a simple bias where we place half the available 
interpolation nodes within the first $1/16$nd of the available support; 
in this case, within 7.5$'$. We found this bias to be a good 
compromise between the need to carefully interpolate the innermost 
portions of the kernel and the need to maintain a sufficient 
number of interpolation points throughout the rest of the kernel.  

\begin{figure} 
  \centering 
  {\includegraphics[width=\columnwidth]{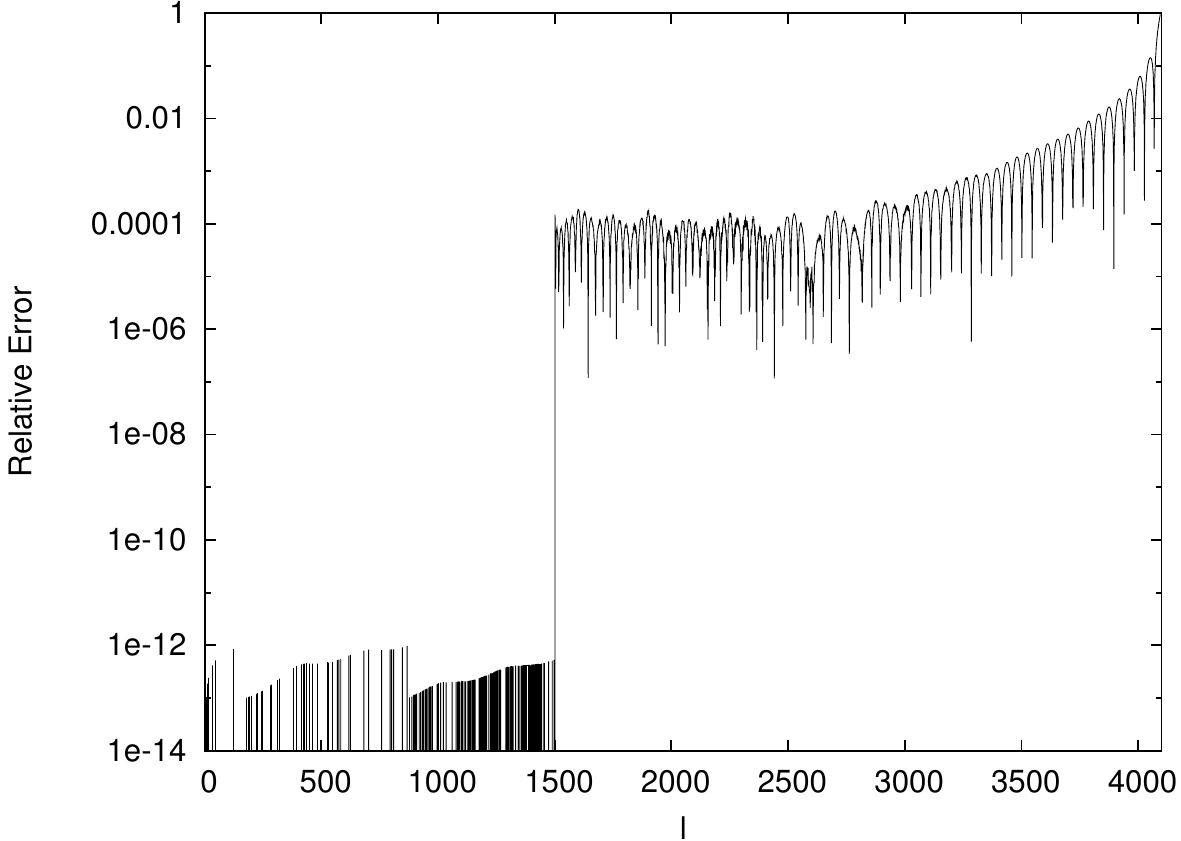}}
  \caption{Estimated relative error of the 
           example approximate kernel $\widetilde{K}_\ell$ to 
           the full kernel $K_\ell$. Shown is the relative error as 
           a function of $\ell$ (Eq.~\ref{eq:relerror}).
           For this example, the 
           $\ell$-space kernel is truncated to $\lcut=1500$ and the
           real-space kernel to $\thetacut=240'$. }
\label{fig:relerror}
\end{figure}

The approximate kernel faithfully represents the full input kernel 
below the truncation threshold of the $\ell$-space kernel at 
$\ell=1500$, which we see in Figure~\ref{fig:relerror}. 
In this figure we show the relative error, defined as
\begin{equation}
  \sigma_\ell = \log_{10} \left| 1 - 
                 \frac{\widetilde{K}_\ell}{K_\ell} \right|.
\label{eq:relerror}
\end{equation}
In this figure we see three distinct regimes. The first, from $\ell=$0-1500 
where the $\ell$-space kernel dominates, has essentially zero error.
From $\ell=$1500 to roughly 3000, we maintain a relative error of roughly 
$10^{-5}$. In this region the real-space kernel is best able to 
reproduce the full input kernel. Finally, at the highest $\ell$, the 
real-space kernel has difficulty following the input kernel and the errors 
begin to exponentially diverge, reaching $100 \%$ 
relative error at $\lmax=4096$. However,
the beam strongly suppresses the kernel here and the high-magnitude 
low-$\ell$ portion
dominates our error estimate. 
Therefore we can ultimately satisfy a given overall error bound.

\begin{figure} 
  \centering 
  {\includegraphics[width=\columnwidth]{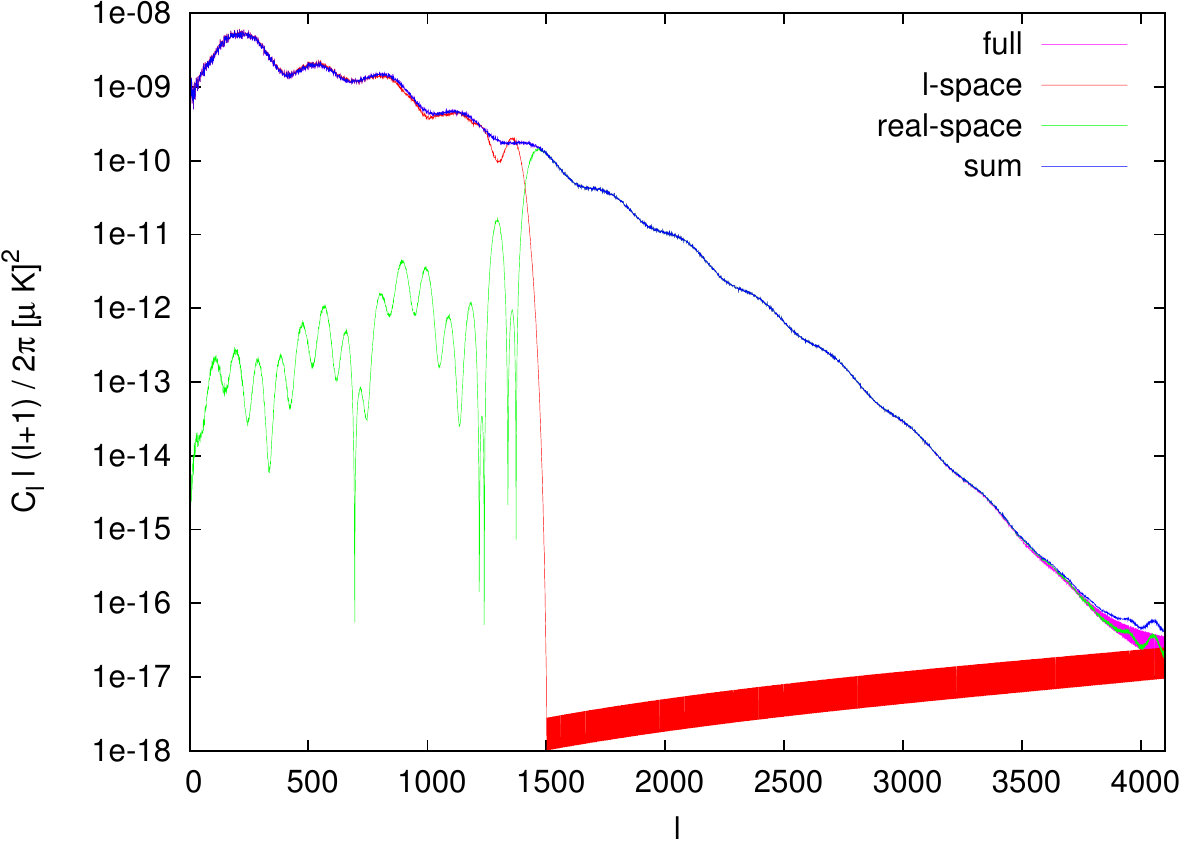}}
  \caption{Derived power spectra after convolving a uniform-noise map
           with various kernels.
           The kernels used are: 
           the full $\ell$-space kernel $K_\ell$ (pink),
           the truncated $\ell$-space kernel $\widehat{K}_\ell$ (red),
           and the truncated real-space kernel $\widehat{K}_\theta$ (green).
           The blue line shows the power spectrum of the map
           created by summing the individual maps of 
           the two truncated kernels.
           For this example, 
           the $\ell$-space kernel is truncated to $\lcut=1500$ and the
           real-space kernel to $\thetacut=240'$. 
}
\label{fig:spectra}
\end{figure}

To evaluate the actual performance of each kernel, we applied them to 
a uniform-noise input map and extracted the spectra.
We compare these spectra in Figure~\ref{fig:spectra}. We show the 
power spectrum after convolving with the full $\ell$-space kernel $K_\ell$,
the truncated $\ell$-space kernel $\widehat{K}_\ell$,
and the truncated real-space kernel $\widehat{K}_\theta$. We also show 
the power spectrum of the summed map. We see that we are able to recover 
the desired power spectrum using the truncated kernels, except at the 
highest $\ell$ range, where interpolation errors and the limitations of 
single-precision arithmetic in the GPU introduce deviations.

\begin{figure} 
  \centering 
  {\includegraphics[width=\columnwidth]{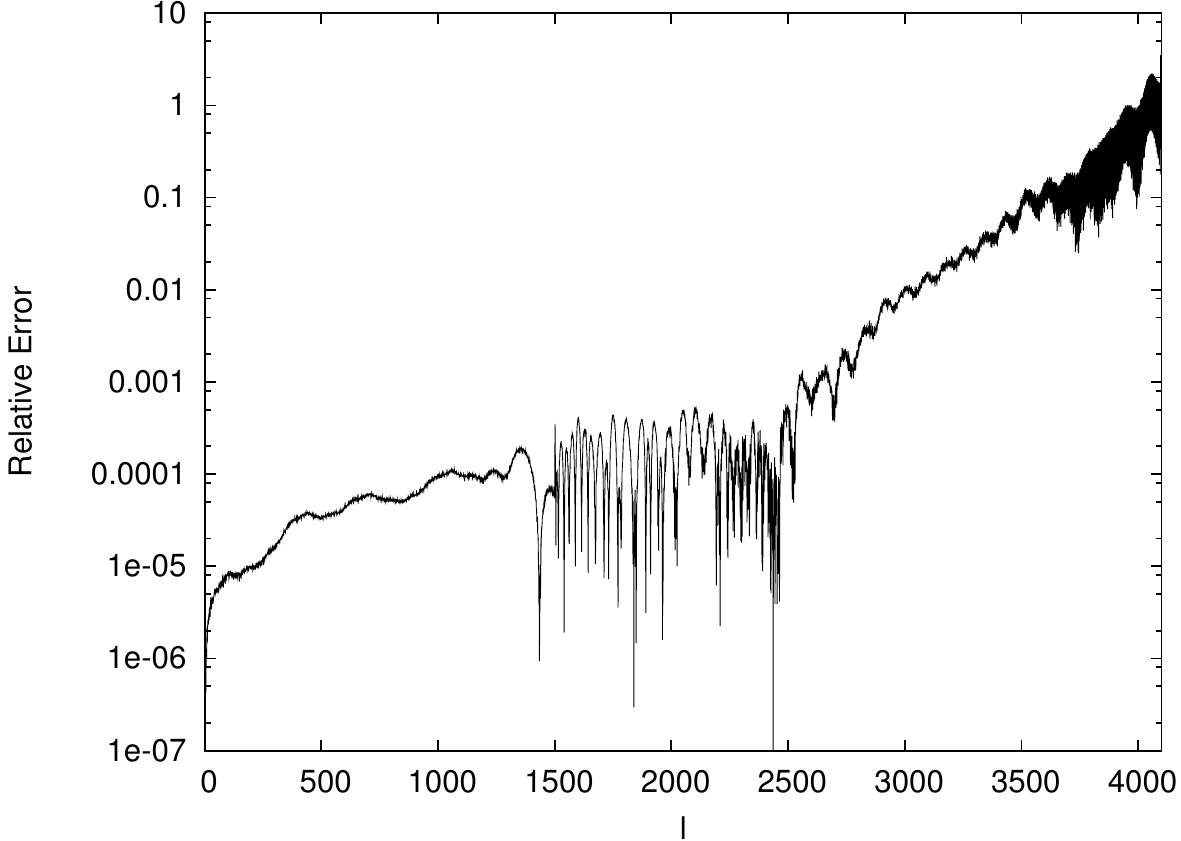}}
  \caption{Actual relative error of the 
           approximate kernel $\widetilde{K}_\ell$ to 
           the full kernel $K_\ell$ after convolution. 
           Shown is the relative error as 
           a function of $\ell$ (Eq.~\ref{eq:relerror}).
           For this example, 
           the $\ell$-space kernel is truncated to $\lcut=1500$ and the
           real-space kernel to $\thetacut=240'$. 
          }
\label{fig:spectrumerror}
\end{figure}

Figure~\ref{fig:spectrumerror} shows the relative error between the 
power spectrum obtained by summing the maps produced by the truncated 
kernels and spectrum obtained by using the full $\ell$-space kernel.
We see similar structure to the estimated relative error, but in this 
case the errors are not negligible below $\lcut=1500$. Here, the difficulty 
of adding the small component due to the real-space kernel to the 
$\ell$-space kernel is apparent. After $\ell=1500$ we see 
small oscillations around the full power spectrum followed by the 
expected exponential rise in the relative error. Altogether, 
we found the total error to be a factor of five higher than estimated 
due to these numerical effects. 
Thus we set the constant $\alpha$ in Eq.(\ref{eq:error}) to five.

\begin{figure*} 
  \centering 
  \subfigure[full kernel]{
    \includegraphics[scale=0.3]{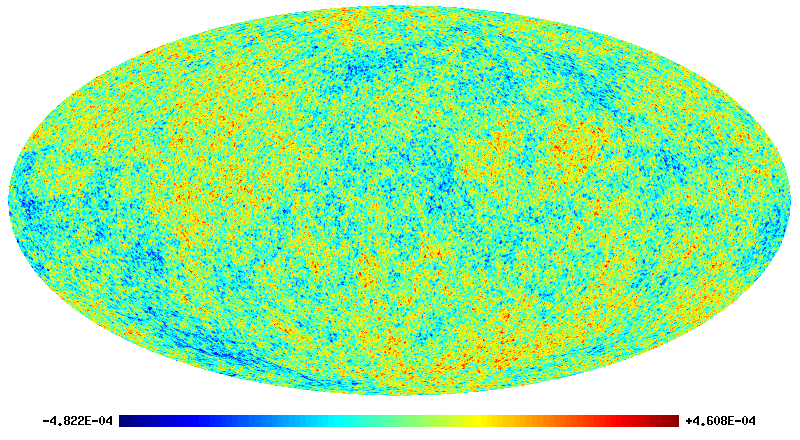}
  }
  \subfigure[residual]{
    \includegraphics[scale=0.3]{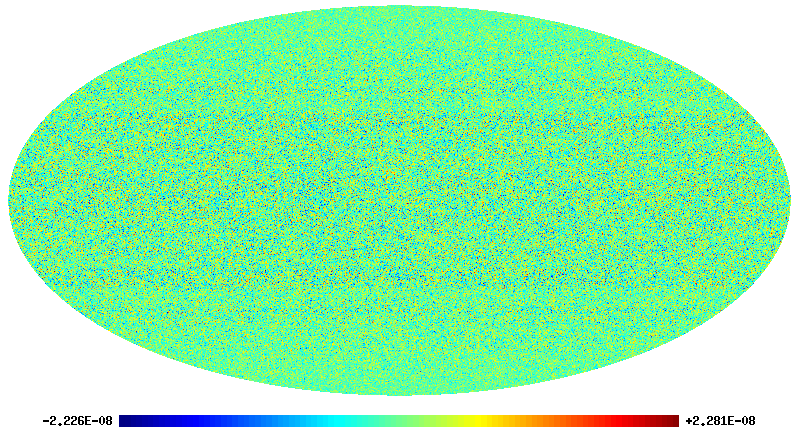}
  }
  \caption{(a) Map after convolving a uniform-noise input map with 
           the full $\ell$-space kernel $K_\ell$.
           (b) The residual between the map
           in panel (a) and the map constructed by summing 
           the convolution outputs of  
           the truncated $\ell$-space kernel $\widehat{K}_\ell$
           and the truncated real-space kernel $\widehat{K}_\theta$.
           For this example, 
           the $\ell$-space kernel is truncated to $\lcut=1500$ and the
           real-space kernel to $\thetacut=240'$.}
\label{fig:maps}
\end{figure*}

\begin{figure*} 
  \centering 
  \subfigure[full kernel]{
    \includegraphics[scale=0.25]{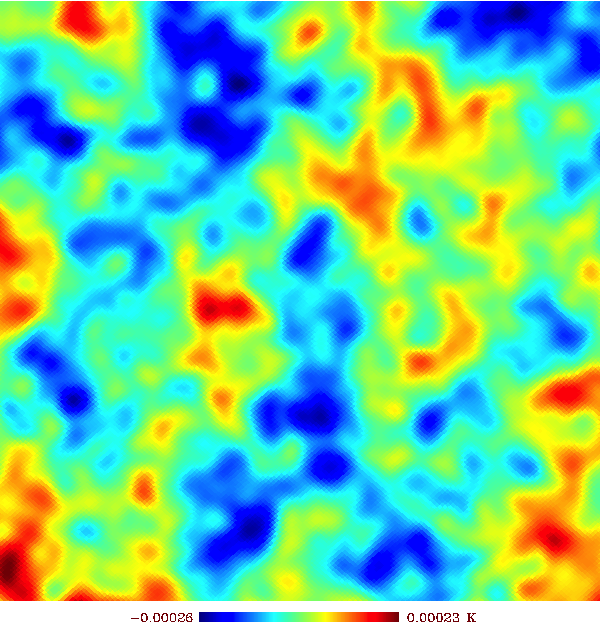}
  }
  \subfigure[$\ell$-space kernel residual]{
    \includegraphics[scale=0.25]{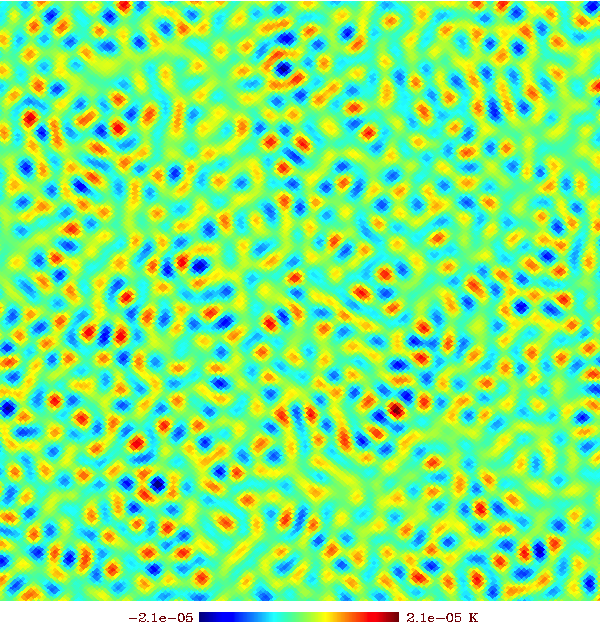}
  }
  \subfigure[real-space kernel]{
    \includegraphics[scale=0.25]{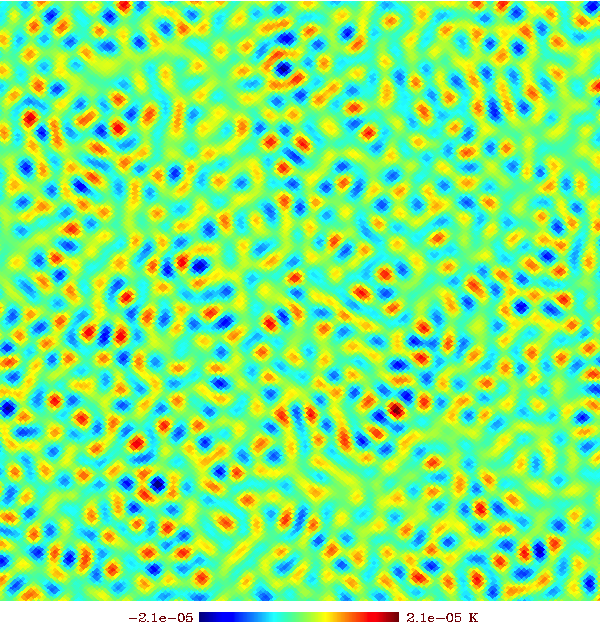}
  }
  \caption{(a) Five-degree patch of the map in Figure~\ref{fig:maps}a. 
           (b) Residual between the map in panel (a) and the 
               map produced by convolving with the 
               the truncated $\ell$-space kernel $\widehat{K}_\ell$.
           (c) Map created by convolving with 
               the truncated real-space kernel $\widehat{K}_\theta$.
           Convolving with the $\ell$-space kernel leaves a small-scale 
           residual that is accurately corrected for with the real-space 
           kernel.
           For this example, 
           the $\ell$-space kernel is truncated to $\lcut=1500$ and the
           real-space kernel to $\thetacut=240'$.}
\label{fig:smallmaps}
\end{figure*}

In Figure~\ref{fig:maps} we show the map after convolving with the 
full $\ell$-space kernel. We also show the residual between this 
map and sum of the maps produced by convolution with the truncated 
$\ell$-space and real-space kernels. We maintain small errors throughout 
the entire map, with the largest errors at the smallest scales, 
as expected.
In Figure~\ref{fig:smallmaps} we show a 5-degree patch of the same 
maps. We see that the 
$\ell$-space kernel reproduces the full map to percent-level accuracy.
However, the real-space kernel is necessary to correctly construct the 
small-scale power and reduce the error to acceptable limits. 

\begin{figure} 
  \centering 
  {\includegraphics[width=\columnwidth]{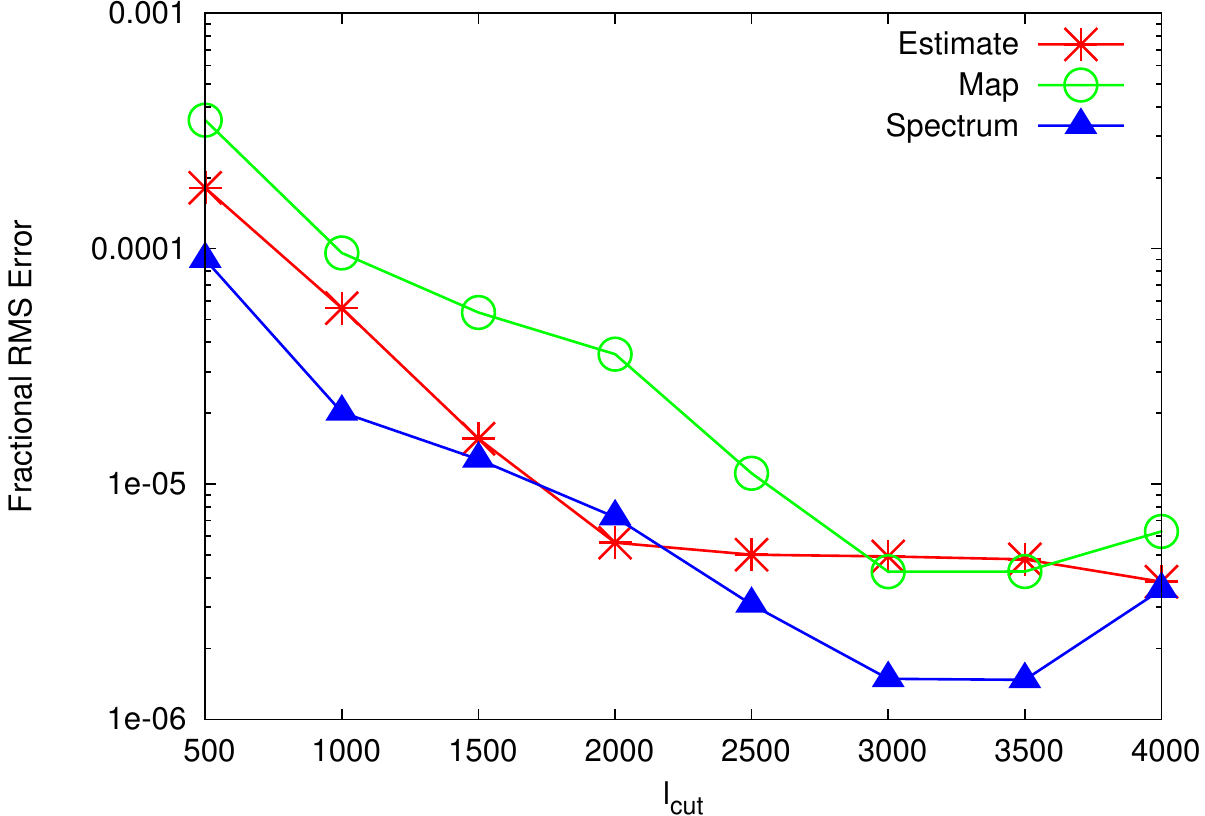}}
  \caption{Estimated 
           RMS error computed by {\tt scytale} (Eq.~\ref{eq:error}) with
           $\alpha=5$ (red stars)
           versus actual RMS error in the maps produced by convolution 
           with a uniform-noise map (green circles) and the RMS error 
           in the power spectra 
           derived from those maps (blue triangles). The lines connecting 
           the points do not 
           represent data but are shown as visual aids.}
\label{fig:error}
\end{figure}

We compare our estimated RMS error to the actual 
map and power spectra errors 
in Figure~\ref{fig:error} for a selection of $\lcut$ values with a
fixed $\thetacut=240'$ and the same 7$'$ beam that we have 
thus far used. For this plot, we have set the empirically-determined 
constant $\alpha$ to five. With this chosen constant, our error 
estimate matches the actual error in the power spectra until 
an $\lcut$ of 2500. At higher $\lcut$ values, we overestimate the 
spectrum errors, but since this lies below our chosen 
error bound of $10^{-5}$ (see below) we choose to maintain this 
value of $\alpha$. The maps tend to produce higher errors, but since 
our quantity of interest is the derived power spectrum, we choose to 
match those errors.

With all this in place
we now turn to our scanning strategy and results of our optimization 
study. We examine beams with 1-10$'$ FWHM, which are most relevant 
to the Planck mission~\citep{MennellaA.2011}. Table~\ref{tab:cuts} shows 
the optimum $(\lcut, \thetacut)$ pairs for five of the ten beam sizes 
studied, assuming a maximum error bound of $10^{-5}$.
Below 6$'$ we could not find suitable truncations that still 
maintained our desired error bound.
We see that all truncations are essentially identical, indicating that 
the ability to split these kernels is binary: either no optimum 
truncations can be found, and that if optimum truncations can be 
found they will be very aggressive. 
For these beam sizes, the optimum $\lcut$ values that 
satisfy the error bounds are significantly below $\lmax$, which promise 
significant enhancements in performance.

\begin{table}
\centering
\caption{Optimum $\lcut$ and $\thetacut$ pairs for each beam FWHM 
         studied, assuming an error bound of $10^{-5}$. }
\begin{tabular}{ccc}
\hline
\hline
Beam FWHM (arcmin) & $\ell_{\rm cut}$ & $\theta_{\rm cut}$ (arcmin) \\
\hline
 7 & 1158 & 390 \\
 8 & 1070 & 390 \\
 9 & 1055 & 360 \\
10 & 979 & 360 \\
11 & 1014 & 330 \\
12 & 960 & 330 \\
13 & 940 & 330 \\
14 & 929 & 300 \\
15 & 961 & 270 \\
\hline
\end{tabular}
\label{tab:cuts}
\end{table}

\begin{figure} 
  \centering 
  {\includegraphics[width=\columnwidth]{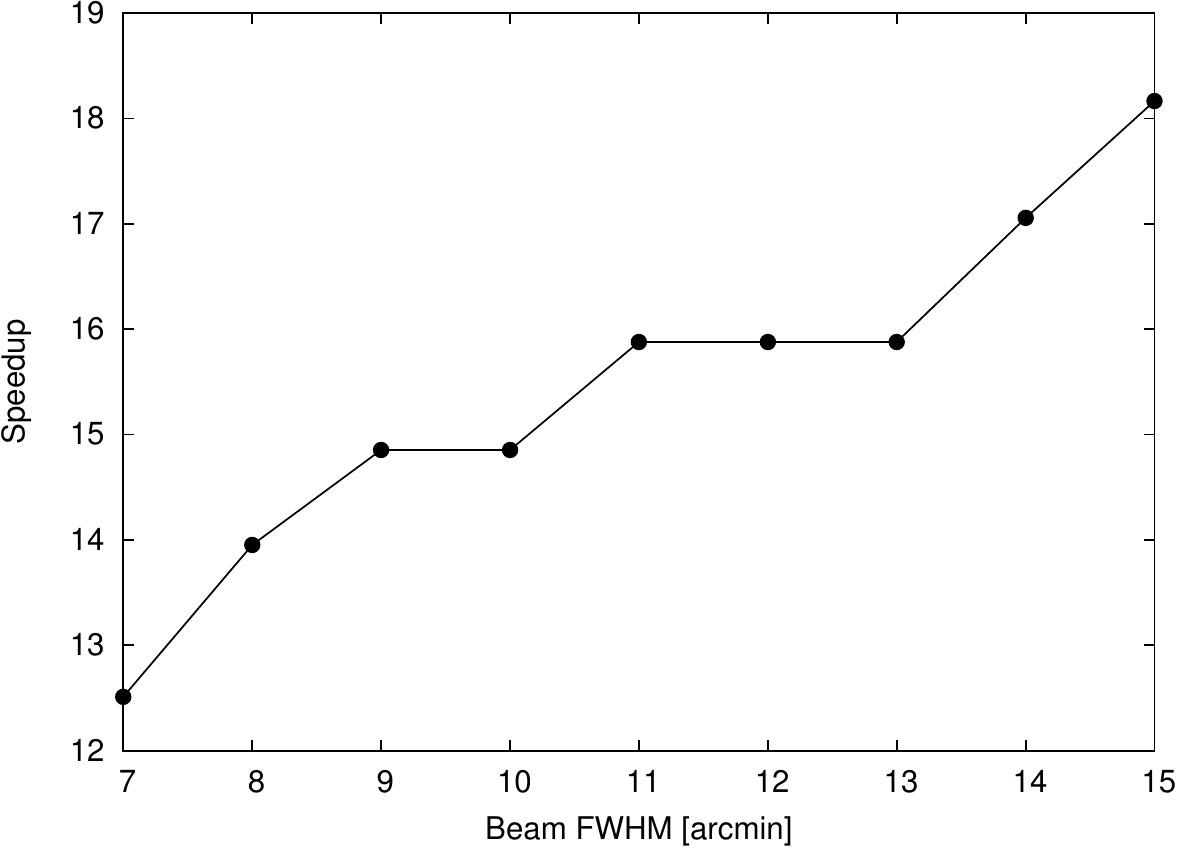}}
  \caption{Speedup versus beam FWHM assuming an overall error bound 
           of $10^{-5}$. See Table~\ref{tab:cuts} for the optimum $\thetacut$ 
           and $\lcut$ values associated with each beam FWHM.}
\label{fig:speedup}
\end{figure}

We show in Figure~\ref{fig:speedup} the speedup versus beam FWHM for these 
beam sizes and our error bound of $10^{-5}$. 
We define the scaling as the time to solution 
with our split approach relative to the cost of applying the
entire kernel (i.e., up to $\lmax$) on the 
CPU with {\tt libpsht}. Below $7'$, we find no optimum truncations 
and hence do not show them. We see significant performance gains
 above $7'$, with the speedups plateauing in the range 12-15. This 
speedup implies a reduction in the computational time from 160 seconds to 
approximately 12 seconds for a single convolution operation. Since all the 
truncations are essentially the same above 7$'$, we find nearly 
identical speedups regardless of the beam size.

\section{Conclusions}
\label{sec:conclusions}

We have introduced and 
discussed a method for splitting radially-symmetric kernels into 
truncated real- and Fourier-space components and estimating the errors 
associated with such splitting. 
We have validated our error estimation 
by performing convolutions with the truncated kernels and computing 
the actual resulting error. 
We have found that for Planck-sized
data sets, a large range of kernels can be split into 
significantly truncated portions while still maintaining an acceptable 
($\sim 10^{-5}$) error bound, leading to significant speedups.

Our analysis was focused on an ideal case; i.e., situations where 
there is no noise and where the input power spectrum remains 
flat. This is the worst-case scenario.
In the case where noise dominates the high-$\ell$ regime we found 
speedups of order $\sim 20$, since we could relax the criterion of 
strictly matching the structure of the full kernel in this region.  

Our approach is currently limited to $\ell \sim 4000$ due to the 
finite amount of fixed memory available on single current-generation GPUs. 
An all-sky convolution up to $\ell=8000$ or $16000$ would require 
splitting the problem across multiple GPUs, as discussed below. However, current experiments that probe 
this regime, such as ACT \citep{KOSOWSKY2003} and 
SPT~\citep{Ruhl2004}, only map on the order of hundreds of square degrees.
By re-orienting their survey maps onto the polar cap, we can keep the 
number HEALPix rings small and exploit our algorithm with currently-available 
GPUs.

While we have focused our analysis on the combination of a single 
GPU working in parallel with a single CPU core, many other 
configurations are possible and indeed can lead to significant 
performance enhancements. For example, the compactness of our 
truncated real-space kernel allows the sky map to be divided 
into latitude bands with minimal overlap. This provides a degree of 
parallelism impossible with general kernels with broad support. 
The scaling in this case is nearly ideal: convolution on the 
latitude bands is completely independent once the necessary overlap 
is included. This scaling holds until the width of the latitude bands 
is equal to twice the kernel width. For example, the truncated real-space 
kernels discussed above, with 
$\thetacut \sim 6^{\circ}$, could potentially be split amongst 30 GPUs. 
The only additional overhead is the cost of 
communicating the overlapping portions. However, since parallel 
GPUs would presumably have independent communication buses, the 
overall communication time would remain relatively constant.
Even if this were not the case the additional communication 
cost could easily be incorporated into our optimization procedure. 

The {\tt ARKCoS} code also has a CPU-based implementation, allowing 
our approach to work on homogeneous architectures. While the speedups 
in the CPU-only case are not as significant, we can still take advantage of the 
parallelism offered by the compact real-space kernels. In this 
scenario, the truncated $\ell$-space kernel can be convolved 
using traditional parallel spherical harmonic transform operations 
on a few cores 
(such as {\tt ccSHT} \footnote{\url{http://crd-legacy.lbl.gov/~cmc/ccSHTlib/doc/index.html}}), 
where the parallel scalability is strongest, while 
the truncated real-space kernel can be convolved using many cores 
in parallel in the manner described above. 
 
Kernel splitting enables the efficient 
allocation of resources for tackling large data sets; in our case, by 
applying real-space kernels with a GPU and 
$\ell$-space kernels with a CPU. 
We have applied this kernel splitting scheme to an optimization study to 
find the realistic speedups associated with splitting a kernel between 
a compact portion to be solved on a GPU and the remainder on a CPU. 
Applying this to kernels and data sets appropriate for the Planck mission, 
we find that this splitting technique can lead to over a factor of 
ten speedup compared to traditional fully CPU-based approaches. 
This significantly improves the feasibility of many necessary and 
important data analysis operations, such as point 
source removal, map making, and power spectrum estimation. 

\section*{Acknowledgments}

The authors acknowledge support from NSF Grant AST-0908902.
This material is based upon work supported in part by NSF Grant AST-1066293 and the hospitality of the Aspen Center for Physics.
PMS is supported by the INFN IS PD51 ``Indark''. This work made in the ILP LABEX (under reference ANR-10-LABX-63) was supported by French state funds managed by the ANR within the Investissements d'Avenir programme under reference ANR-11-IDEX-0004-02.
FE was partially supported by a New Frontiers in Astronomy and Cosmology grant \#37426, and the European Research Council under the European Community's Seventh Framework Programme (FP7/2007-2013) / ERC grant agreement no 306478-CosmicDawn.

\bibliographystyle{model2-names}
\bibliography{scytale}		

\end{document}